\documentclass[     ,final              ]{aipproc}%
\usepackage{amssymb}
\usepackage{amsmath}
\usepackage{amsfonts}
\usepackage{graphicx}%
\layoutstyle{6x9}
\begin{document}

\title{Modified Dispersion Relations and Noncommutative Geometry lead to a finite
Zero Point Energy}

\classification{ 04.60.-m, 04.62+v, 05.10.Cc}
\keywords{Cosmological Constant, Quantum Cosmology, Quantum Gravity, Noncommutative Geometry, Modified Dispersion Relations}
\author{Remo Garattini}{address={Universit\`{a} degli Studi di Bergamo, Facolt\`{a} di Ingegneria,
\\ Viale Marconi 5, 24044 Dalmine (Bergamo) Italy and\\ I.N.F.N. -
sezione di Milano, Milan, Italy.\\ E-mail:remo.garattini@unibg.it}}

\begin{abstract}
We compute Zero Point Energy in a spherically symmetric background with the
help of the Wheeler-DeWitt equation. This last one is regarded as a
Sturm-Liouville problem with the cosmological constant considered as the
associated eigenvalue. The graviton contribution, at one loop is extracted wit
the help of a variational approach together with Gaussian trial functionals.
The divergences handled with a zeta function regularization are compared with
the results obtained using a Noncommutative Geometry (NCG) and Modified
Dispersion Relations (MDR). In both NCG and MDR no renormalization scheme is
necessary to remove infinities in contrast to what happens in conventional approaches.

\end{abstract}
\maketitle

\section{Introduction}

Quantum Gravity should be the fusion between Quantum Mechanics and General
Relativity. When they are disconnected these theories work pretty good in
their respective domains. However, when one tries to merge them in a single
theory, one discovers that this last one is non renormalizable and therefore
its predictive power is lost. Recently, Ho\v{r}ava proposed a modification of
Einstein gravity motivated by the Lifshitz theory in solid state
physics\cite{Horava}\cite{Lifshitz}. Such modification allows the theory to be
power-counting ultraviolet (UV)-renormalizable and should recover general
relativity in the infrared (IR) limit. Nevertheless Ho\v{r}ava-Lifshitz (HL)
theory is non-covariant, at least in its original formulation\footnote{See
Ref.\cite{Horava1}. for recent progress to obtain a covariant theory.}.
Indeed, in this approach space and time exhibit Lifshitz scale invariance of
the form%
\begin{equation}
t\rightarrow\ell^{z}t\ \text{\textrm{and}}\ x^{i}\rightarrow\ell x^{i}%
\end{equation}
with $z\geq1$. $z$ is called the dynamical critical exponent. The breaking of
the 4D diffeomorphism invariance allows a different treatment of the kinetic
and potential terms for the metric: from one side the kinetic term is
quadratic in time derivatives of the metric, from the other side the potential
has high-order space derivatives. In particular the UV behavior is dominated
by the square of the Cotton tensor of the $3D$ geometry by means of a $k^{6}$
contribution to the propagator leading to a renormalizable power-counting
theory. The original HL\ theory is based on two assumptions -- detailed
balance and projectability\cite{Horava2}. The projectability condition is a
weak version of the invariance with respect to time reparametrization and
therefore to the Wheeler-DeWitt (WDW) equation\cite{DeWitt}. In this
contribution, we do not want to discuss the appealing properties of the HL
theory, but rather, we would like to investigate the original WDW equation
with a look to possible deformations induced by the spacetime itself.
Therefore no matter fields will be included in such a discussion. We have in
mind two interesting cases: Non Commutative Geometry and Rainbow's Gravity.
Both of these cases are modifications of the original space time structure.
For this reason we expect some interesting consequences on the computation of
some observables. In this paper, we are interested to see what are the effects
of such distortions on Zero Point Energy (ZPE). A partial promising answer has
been obtained in Ref.\cite{RGPLB}, where the UV divergences ogf a
Schwarzschild black hole are kept under control by an appropriate choice of
the Rainbow's metric. Units in which $\hbar=c=k=1$ are used throughout the paper.

\section{The WDW Equation}

The WDW equation was originally introduced by Bryce DeWitt as an attempt to
quantize General Relativity in a Hamiltonian formulation. It is described by%
\begin{equation}
\mathcal{H}\Psi=\left[  \left(  2\kappa\right)  G_{ijkl}\pi^{ij}\pi^{kl}%
-\frac{\sqrt{g}}{2\kappa}\!{}\!\left(  \,\!^{3}R-2\Lambda\right)  \right]
\Psi=0\label{WDW}%
\end{equation}
and it represents the quantum version of the classical constraint which
guarantees the invariance under time reparametrization. $G_{ijkl}$ is the
super-metric, $\pi^{ij}$ is the super-momentum,$^{3}R$ is the scalar curvature
in three dimensions and $\Lambda$ is the cosmological constant, while
$\kappa=8\pi G$ with $G$ the Newton's constant. In this way, the WDW equation
is written in its most general form. The main reason to use such an equation
to discuss renormalization problems is related to the possibility of formally
re-writing the WDW equation as an expectation value computation\footnote{See
also Ref.\cite{CG:2007} for an application of the method to a $f\left(
R\right)  $ theory.}\cite{Remo}. Indeed, if we multiply Eq.$\left(
\ref{WDW}\right)  $ by $\Psi^{\ast}\left[  g_{ij}\right]  $ and functionally
integrate over the three spatial metric $g_{ij}$ we find%
\begin{equation}
\frac{1}{V}\frac{\int\mathcal{D}\left[  g_{ij}\right]  \Psi^{\ast}\left[
g_{ij}\right]  \int_{\Sigma}d^{3}x\hat{\Lambda}_{\Sigma}\Psi\left[
g_{ij}\right]  }{\int\mathcal{D}\left[  g_{ij}\right]  \Psi^{\ast}\left[
g_{ij}\right]  \Psi\left[  g_{ij}\right]  }=\frac{1}{V}\frac{\left\langle
\Psi\left\vert \int_{\Sigma}d^{3}x\hat{\Lambda}_{\Sigma}\right\vert
\Psi\right\rangle }{\left\langle \Psi|\Psi\right\rangle }=-\frac{\Lambda
}{\kappa}.\label{VEV}%
\end{equation}
In Eq.$\left(  \ref{VEV}\right)  $ we have also integrated over the
hypersurface $\Sigma$ and we have defined%
\begin{equation}
V=\int_{\Sigma}d^{3}x\sqrt{g}%
\end{equation}
as the volume of the hypersurface $\Sigma$ with%
\begin{equation}
\hat{\Lambda}_{\Sigma}=\left(  2\kappa\right)  G_{ijkl}\pi^{ij}\pi^{kl}%
-\sqrt{g}^{3}R/\left(  2\kappa\right)  .\label{LambdaSigma}%
\end{equation}
In this form, Eq.$\left(  \ref{VEV}\right)  $ can be used to compute ZPE
provided that $\Lambda/\kappa$ be considered as an eigenvalue of $\hat
{\Lambda}_{\Sigma}$. In particular, Eq.$\left(  \ref{VEV}\right)  $ represents
the Sturm-Liouville problem associated with the cosmological constant. To
solve Eq.$\left(  \ref{VEV}\right)  $ is a quite impossible task. Therefore,
we are oriented to use a variational approach with trial wave functionals. The
related boundary conditions are dictated by the choice of the trial wave
functionals which, in our case are of the Gaussian type. Different types of
wave functionals correspond to different boundary conditions. The choice of a
Gaussian wave functional is justified by the fact that ZPE should be described
by a good candidate of the \textquotedblleft\textit{vacuum state}%
\textquotedblright. To fix ideas, we choose the following form of the metric%
\begin{equation}
ds^{2}=-N^{2}\left(  r\right)  dt^{2}+\frac{dr^{2}}{1-\frac{b\left(  r\right)
}{r}}+r^{2}\left(  d\theta^{2}+\sin^{2}\theta d\phi^{2}\right)  ,\label{dS}%
\end{equation}
where $b\left(  r\right)  $ is subject to the only condition $b\left(
r_{t}\right)  =r_{t}$. As a first step, we decompose the gravitational
perturbation in such a way to obtain the graviton contribution enclosed in
Eq.$\left(  \ref{VEV}\right)  $.

\subsection{Extracting the graviton contribution}

\label{ps1}We can gain more information if we consider $g_{ij}=\bar{g}%
_{ij}+h_{ij},$where $\bar{g}_{ij}$ is the background metric and $h_{ij}$ is a
quantum fluctuation around the background. Thus Eq.$\left(  \ref{VEV}\right)
$ can be expanded in terms of $h_{ij}$. Since the kinetic part of
$\hat{\Lambda}_{\Sigma}$ is quadratic in the momenta, we only need to expand
the three-scalar curvature $\int d^{3}x\sqrt{g}{}^{3}R$ up to the quadratic
order. However, to proceed with the computation, we also need an orthogonal
decomposition on the tangent space of 3-metric
deformations\cite{Vassilevich:1993,Quad:1969}:%

\begin{equation}
h_{ij}=\frac{1}{3}\left(  \sigma+2\nabla\cdot\xi\right)  g_{ij}+\left(
L\xi\right)  _{ij}+h_{ij}^{\bot}. \label{p21a}%
\end{equation}
The operator $L$ maps $\xi_{i}$ into symmetric tracefree tensors%
\begin{equation}
\left(  L\xi\right)  _{ij}=\nabla_{i}\xi_{j}+\nabla_{j}\xi_{i}-\frac{2}%
{3}g_{ij}\left(  \nabla\cdot\xi\right)  ,
\end{equation}
$h_{ij}^{\bot}$ is the traceless-transverse component of the perturbation
(TT), namely $g^{ij}h_{ij}^{\bot}=0$, $\nabla^{i}h_{ij}^{\bot}=0$ and $h$ is
the trace of $h_{ij}$. It is immediate to recognize that the trace element
$\sigma=h-2\left(  \nabla\cdot\xi\right)  $ is gauge invariant. If we perform
the same decomposition also on the momentum $\pi^{ij}$, up to second order
Eq.$\left(  \ref{VEV}\right)  $ becomes%
\begin{equation}
\frac{1}{V}\frac{\left\langle \Psi\left\vert \int_{\Sigma}d^{3}x\left[
\hat{\Lambda}_{\Sigma}^{\bot}+\hat{\Lambda}_{\Sigma}^{\xi}+\hat{\Lambda
}_{\Sigma}^{\sigma}\right]  ^{\left(  2\right)  }\right\vert \Psi\right\rangle
}{\left\langle \Psi|\Psi\right\rangle }=-\frac{\Lambda}{\kappa}.
\label{lambda0_2}%
\end{equation}
Concerning the measure appearing in Eq.$\left(  \ref{VEV}\right)  $, we have
to note that the decomposition $\left(  \ref{p21a}\right)  $ induces the
following transformation on the functional measure $\mathcal{D}h_{ij}%
\rightarrow\mathcal{D}h_{ij}^{\bot}\mathcal{D}\xi_{i}\mathcal{D}\sigma J_{1}$,
where the Jacobian related to the gauge vector variable $\xi_{i}$ is%
\begin{equation}
J_{1}=\left[  \det\left(  \bigtriangleup g^{ij}+\frac{1}{3}\nabla^{i}%
\nabla^{j}-R^{ij}\right)  \right]  ^{\frac{1}{2}}.
\end{equation}
This is nothing but the famous Faddev-Popov determinant. It becomes more
transparent if $\xi_{a}$ is further decomposed into a transverse part $\xi
_{a}^{T}$ with $\nabla^{a}\xi_{a}^{T}=0$ and a longitudinal part $\xi
_{a}^{\parallel}$ with $\xi_{a}^{\parallel}=$ $\nabla_{a}\psi$, then $J_{1}$
can be expressed by an upper triangular matrix for certain backgrounds (e.g.
Schwarzschild in three dimensions). It is immediate to recognize that for an
Einstein space in any dimension, cross terms vanish and $J_{1}$ can be
expressed by a block diagonal matrix. Since $\det AB=\det A\det B$, the
functional measure $\mathcal{D}h_{ij}$ factorizes into%
\begin{equation}
\mathcal{D}h_{ij}=\left(  \det\bigtriangleup_{V}^{T}\right)  ^{\frac{1}{2}%
}\left(  \det\left[  \frac{2}{3}\bigtriangleup^{2}+\nabla_{i}R^{ij}\nabla
_{j}\right]  \right)  ^{\frac{1}{2}}\mathcal{D}h_{ij}^{\bot}\mathcal{D}\xi
^{T}\mathcal{D}\psi
\end{equation}
with $\left(  \bigtriangleup_{V}^{ij}\right)  ^{T}=\bigtriangleup
g^{ij}-R^{ij}$ acting on transverse vectors, which is the Faddeev-Popov
determinant. In writing the functional measure $\mathcal{D}h_{ij}$, we have
here ignored the appearance of a multiplicative anomaly\cite{EVZ:1998}. Thus
the inner product can be written as%
\begin{equation}
\int\mathcal{D\rho}\Psi^{\ast}\left[  h_{ij}^{\bot}\right]  \Psi^{\ast}\left[
\xi^{T}\right]  \Psi^{\ast}\left[  \sigma\right]  \Psi\left[  h_{ij}^{\bot
}\right]  \Psi\left[  \xi^{T}\right]  \Psi\left[  \sigma\right]  ,
\end{equation}
where%
\begin{equation}
\mathcal{D}\rho=\mathcal{D}h_{ij}^{\bot}\mathcal{D}\xi^{T}\mathcal{D}%
\sigma\left(  \det\bigtriangleup_{V}^{T}\right)  ^{\frac{1}{2}}\left(
\det\left[  \frac{2}{3}\bigtriangleup^{2}+\nabla_{i}R^{ij}\nabla_{j}\right]
\right)  ^{\frac{1}{2}}.
\end{equation}
Nevertheless, since there is no interaction between ghost fields and the other
components of the perturbation at this level of approximation, the Jacobian
appearing in the numerator and in the denominator simplify. The reason can be
found in terms of connected and disconnected terms. The disconnected terms
appear in the Faddeev-Popov determinant and these ones are not linked by the
Gaussian integration. This means that disconnected terms in the numerator and
the same ones appearing in the denominator cancel out. Therefore, Eq.$\left(
\ref{lambda0_2}\right)  $ factorizes into three pieces. The piece containing
$\hat{\Lambda}_{\Sigma}^{\bot}$ is the contribution of the
transverse-traceless tensors (TT): essentially is the graviton contribution
representing true physical degrees of freedom. Regarding the vector term
$\hat{\Lambda}_{\Sigma}^{T}$, we observe that under the action of
infinitesimal diffeomorphism generated by a vector field $\epsilon_{i}$, the
components of $\left(  \ref{p21a}\right)  $ transform as
follows\cite{Vassilevich:1993}%
\begin{equation}
\xi_{j}\longrightarrow\xi_{j}+\epsilon_{j},\qquad h\longrightarrow
h+2\nabla\cdot\xi,\qquad h_{ij}^{\bot}\longrightarrow h_{ij}^{\bot}.
\end{equation}
The Killing vectors satisfying the condition $\nabla_{i}\xi_{j}+\nabla_{j}%
\xi_{i}=0,$ do not change $h_{ij}$, and thus should be excluded from the gauge
group. All other diffeomorphisms act on $h_{ij}$ nontrivially. We need to fix
the residual gauge freedom on the vector $\xi_{i}$. The simplest choice is
$\xi_{i}=0$. This new gauge fixing produces the same Faddeev-Popov determinant
connected to the Jacobian $J_{1}$ and therefore will not contribute to the
final value. We are left with%
\begin{equation}
\frac{1}{V}\frac{\left\langle \Psi^{\bot}\left\vert \int_{\Sigma}d^{3}x\left[
\hat{\Lambda}_{\Sigma}^{\bot}\right]  ^{\left(  2\right)  }\right\vert
\Psi^{\bot}\right\rangle }{\left\langle \Psi^{\bot}|\Psi^{\bot}\right\rangle
}+\frac{1}{V}\frac{\left\langle \Psi^{\sigma}\left\vert \int_{\Sigma}%
d^{3}x\left[  \hat{\Lambda}_{\Sigma}^{\sigma}\right]  ^{\left(  2\right)
}\right\vert \Psi^{\sigma}\right\rangle }{\left\langle \Psi^{\sigma}%
|\Psi^{\sigma}\right\rangle }=-\frac{\Lambda^{\bot}}{\kappa}-\frac
{\Lambda^{\sigma}}{\kappa}. \label{lambda0_2a}%
\end{equation}
Note that in the expansion of $\int_{\Sigma}d^{3}x\sqrt{g}{}R$ to second
order, a coupling term between the TT component and scalar one remains.
However, the Gaussian integration does not allow such a mixing which has to be
introduced with an appropriate wave functional. Extracting the TT tensor
contribution from Eq.$\left(  \ref{VEV}\right)  $ approximated to second order
in perturbation of the spatial part of the metric into a background term
$\bar{g}_{ij}$, and a perturbation $h_{ij}$, we get%
\begin{equation}
\hat{\Lambda}_{\Sigma}^{\bot}=\frac{1}{4V}\int_{\Sigma}d^{3}x\sqrt{\bar{g}%
}G^{ijkl}\left[  \left(  2\kappa\right)  K^{-1\bot}\left(  x,x\right)
_{ijkl}+\frac{1}{\left(  2\kappa\right)  }\!{}\left(  \tilde{\bigtriangleup
}_{L\!}\right)  _{j}^{a}K^{\bot}\left(  x,x\right)  _{iakl}\right]  ,
\label{p22}%
\end{equation}
where%
\begin{equation}
\left(  \tilde{\bigtriangleup}_{L\!}\!{}h^{\bot}\right)  _{ij}=\left(
\bigtriangleup_{L\!}\!{}h^{\bot}\right)  _{ij}-4R{}_{i}^{k}\!{}h_{kj}^{\bot
}+\text{ }^{3}R{}\!{}h_{ij}^{\bot} \label{M Lichn}%
\end{equation}
is the modified Lichnerowicz operator and $\bigtriangleup_{L}$is the
Lichnerowicz operator defined by%
\begin{equation}
\left(  \bigtriangleup_{L}h\right)  _{ij}=\bigtriangleup h_{ij}-2R_{ikjl}%
h^{kl}+R_{ik}h_{j}^{k}+R_{jk}h_{i}^{k}\qquad\bigtriangleup=-\nabla^{a}%
\nabla_{a}. \label{DeltaL}%
\end{equation}
$G^{ijkl}$ represents the inverse DeWitt metric and all indices run from one
to three. Note that the term $-4R{}_{i}^{k}\!{}h_{kj}^{\bot}+$ $^{3}R{}%
\!{}h_{ij}^{\bot}$ disappears in four dimensions. The propagator $K^{\bot
}\left(  x,x\right)  _{iakl}$ can be represented as
\begin{equation}
K^{\bot}\left(  \overrightarrow{x},\overrightarrow{y}\right)  _{iakl}%
=\sum_{\tau}\frac{h_{ia}^{\left(  \tau\right)  \bot}\left(  \overrightarrow
{x}\right)  h_{kl}^{\left(  \tau\right)  \bot}\left(  \overrightarrow
{y}\right)  }{2\lambda\left(  \tau\right)  }, \label{proptt}%
\end{equation}
where $h_{ia}^{\left(  \tau\right)  \bot}\left(  \overrightarrow{x}\right)  $
are the eigenfunctions of $\tilde{\bigtriangleup}_{L\!}$. $\tau$ denotes a
complete set of indices and $\lambda\left(  \tau\right)  $ are a set of
variational parameters to be determined by the minimization of Eq.$\left(
\ref{p22}\right)  $. The expectation value of $\hat{\Lambda}_{\Sigma}^{\bot}$
is easily obtained by inserting the form of the propagator into Eq.$\left(
\ref{p22}\right)  $ and minimizing with respect to the variational function
$\lambda\left(  \tau\right)  $. Thus the total one loop energy density for TT
tensors becomes%
\begin{equation}
\frac{\Lambda}{8\pi G}=-\frac{1}{2}\sum_{\tau}\left[  \sqrt{\omega_{1}%
^{2}\left(  \tau\right)  }+\sqrt{\omega_{2}^{2}\left(  \tau\right)  }\right]
. \label{1loop}%
\end{equation}
The above expression makes sense only for $\omega_{i}^{2}\left(  \tau\right)
>0$, where $\omega_{i}$ are the eigenvalues of $\tilde{\bigtriangleup}_{L\!}$.
In the next section, we will explicitly evaluate Eq.$\left(  \ref{1loop}%
\right)  $ for a background of spherically symmetric type.

\section{One loop energy density: Conventional Regularization and
Renormalization}

The reference metric $\left(  \ref{dS}\right)  $ can be cast into the
following form%
\begin{equation}
ds^{2}=-N^{2}\left(  r\left(  x\right)  \right)  dt^{2}+dx^{2}+r^{2}\left(
x\right)  \left(  d\theta^{2}+\sin^{2}\theta d\phi^{2}\right)  ,
\label{metric}%
\end{equation}
where%
\begin{equation}
dx=\pm\frac{dr}{\sqrt{1-\frac{b\left(  r\right)  }{r}}}. \label{dx}%
\end{equation}
Specific examples are%
\begin{equation}
b\left(  r\right)  =\frac{\Lambda_{dS}}{3}r^{3};\qquad b\left(  r\right)
=-\frac{\Lambda_{AdS}}{3}r^{3}\qquad\mathrm{and}\qquad b\left(  r\right)
=2MG.
\end{equation}
However, we would like to maintain the form of the line element $\left(
\ref{metric}\right)  $ as general as possible. With the help of Regge and
Wheeler representation\cite{Regge Wheeler:1957}, the Lichnerowicz operator
$\left(  \tilde{\bigtriangleup}_{L\!}\!{}h^{\bot}\right)  _{ij}$ can be
reduced to%
\begin{equation}
\left[  -\frac{d^{2}}{dx^{2}}+\frac{l\left(  l+1\right)  }{r^{2}}+m_{i}%
^{2}\left(  r\right)  \right]  f_{i}\left(  x\right)  =\omega_{i,l}^{2}%
f_{i}\left(  x\right)  \quad i=1,2\quad, \label{p34}%
\end{equation}
where we have used reduced fields of the form $f_{i}\left(  x\right)
=F_{i}\left(  x\right)  /r$ and where we have defined two r-dependent
effective masses $m_{1}^{2}\left(  r\right)  $ and $m_{2}^{2}\left(  r\right)
$%
\begin{equation}
\left\{
\begin{array}
[c]{c}%
m_{1}^{2}\left(  r\right)  =\frac{6}{r^{2}}\left(  1-\frac{b\left(  r\right)
}{r}\right)  +\frac{3}{2r^{2}}b^{\prime}\left(  r\right)  -\frac{3}{2r^{3}%
}b\left(  r\right) \\
\\
m_{2}^{2}\left(  r\right)  =\frac{6}{r^{2}}\left(  1-\frac{b\left(  r\right)
}{r}\right)  +\frac{1}{2r^{2}}b^{\prime}\left(  r\right)  +\frac{3}{2r^{3}%
}b\left(  r\right)
\end{array}
\right.  \quad\left(  r\equiv r\left(  x\right)  \right)  . \label{masses}%
\end{equation}
In order to use the W.K.B. method considered by `t Hooft in the brick wall
problem\cite{tHooft:1985}, from Eq.$\left(  \ref{p34}\right)  $ we can extract
two r-dependent radial wave numbers%
\begin{equation}
k_{i}^{2}\left(  r,l,\omega_{i,nl}\right)  =\omega_{i,nl}^{2}-\frac{l\left(
l+1\right)  }{r^{2}}-m_{i}^{2}\left(  r\right)  \quad i=1,2\quad. \label{kTT}%
\end{equation}
Then the counting of the number of modes with frequency less than $\omega_{i}$
is given approximately by%
\begin{equation}
\tilde{g}\left(  \omega_{i}\right)  =\int_{0}^{l_{\max}}\nu_{i}\left(
l,\omega_{i}\right)  \left(  2l+1\right)  dl. \label{p41}%
\end{equation}
$\nu_{i}\left(  l,\omega_{i}\right)  $ is the number of nodes in the mode with
$\left(  l,\omega_{i}\right)  $, such that $\left(  r\equiv r\left(  x\right)
\right)  $
\begin{equation}
\nu_{i}\left(  l,\omega_{i}\right)  =\frac{1}{\pi}\int_{-\infty}^{+\infty
}dx\sqrt{k_{i}^{2}\left(  r,l,\omega_{i}\right)  }. \label{p42}%
\end{equation}
Here it is understood that the integration with respect to $x$ and $l_{\max}$
is taken over those values which satisfy $k_{i}^{2}\left(  r,l,\omega
_{i}\right)  \geq0$. With the help of Eqs.$\left(  \ref{p41},\ref{p42}\right)
$, Eq.$\left(  \ref{1loop}\right)  $ becomes%
\begin{equation}
\frac{\Lambda}{8\pi G}=-\frac{1}{\pi}\sum_{i=1}^{2}\int_{0}^{+\infty}%
\omega_{i}\frac{d\tilde{g}\left(  \omega_{i}\right)  }{d\omega_{i}}d\omega
_{i}. \label{tot1loop}%
\end{equation}
This is the one loop graviton contribution to the induced cosmological
constant. The explicit evaluation of Eq.$\left(  \ref{tot1loop}\right)  $
gives%
\begin{equation}
\frac{\Lambda}{8\pi G}=\rho_{1}+\rho_{2}=-\frac{1}{4\pi^{2}}\sum_{i=1}^{2}%
\int_{\sqrt{m_{i}^{2}\left(  r\right)  }}^{+\infty}\omega_{i}^{2}\sqrt
{\omega_{i}^{2}-m_{i}^{2}\left(  r\right)  }d\omega_{i}, \label{t1l}%
\end{equation}
where we have included an additional $4\pi$ coming from the angular
integration. The use of the zeta function regularization method to compute the
energy densities $\rho_{1}$ and $\rho_{2}$ leads to%
\begin{equation}
\rho_{i}\left(  \varepsilon\right)  =\frac{m_{i}^{4}\left(  r\right)  }%
{64\pi^{2}}\left[  \frac{1}{\varepsilon}+\ln\left(  \frac{4\mu^{2}}{m_{i}%
^{2}\left(  r\right)  \sqrt{e}}\right)  \right]  \quad i=1,2\quad,
\label{rhoe}%
\end{equation}
where we have introduced the additional mass parameter $\mu$ in order to
restore the correct dimension for the regularized quantities. Such an
arbitrary mass scale emerges unavoidably in any regularization scheme. The
renormalization is performed via the absorption of the divergent part into the
re-definition of a bare classical quantity. Here we have two possible choices:
the induced cosmological constant $\Lambda$ or the gravitational Newton
constant $G$. If we decide to absorb the divergence with the help of the
cosmological constant $\Lambda$, we have to separate it into a bare
cosmological constant $\Lambda_{0}$ and a divergent quantity $\Lambda^{div}$,
where%
\begin{equation}
\Lambda^{div}=\frac{Gm_{0}^{4}\left(  r\right)  }{\varepsilon32\pi^{2}},
\end{equation}
and the remaining finite value for the cosmological constant reads%
\begin{equation}
\frac{\Lambda_{0}}{8\pi G}=\left(  \rho_{1}\left(  \mu\right)  +\rho
_{2}\left(  \mu\right)  \right)  =\rho_{eff}^{TT}\left(  \mu,r\right)
=\frac{m_{0}^{4}\left(  r\right)  }{32\pi^{2}}\ln\left(  \frac{4\mu^{2}}%
{m_{0}^{2}\left(  r\right)  \sqrt{e}}\right)  . \label{lambda0}%
\end{equation}

\section{The Example of Non Commutative Theories}

Non Commutative theories provide a powerful method to naturally regularize
divergent integrals appearing in Eq.$\left(  \ref{t1l}\right)  $. Basically,
the number of states is modified in the following way\cite{RG PN}%
\begin{equation}
dn=\frac{d^{3}xd^{3}k}{\left(  2\pi\right)  ^{3}}\ \Longrightarrow
\ dn_{i}=\frac{d^{3}xd^{3}k}{\left(  2\pi\right)  ^{3}}\exp\left(
-\frac{\theta}{4}\left(  \omega_{i,nl}^{2}-m_{i}^{2}\left(  r\right)  \right)
\right)  ,\quad i=1,2. \label{moddn}%
\end{equation}
This deformation corresponds to an effective cut off on the background
geometry $\left(  \ref{metric}\right)  $. The UV cut off is triggered only by
higher momenta modes $\gtrsim1/\sqrt{\theta}$ which propagate over the
background geometry. The virtue of this kind of deformation is its exponential
damping profile, which encodes an intrinsic nonlocal character into fields
$f_{i}(x)$. Plugging $\left(  \ref{p42}\right)  $ into $\left(  \ref{p41}%
\right)  $ and taking account of $\left(  \ref{moddn}\right)  $, the number of
modes with frequency less than $\omega_{i}$, $i=1,2$ is given by%
\begin{equation}
\tilde{g}\left(  \omega_{i}\right)  =\frac{1}{\pi}\int_{-\infty}^{+\infty
}dx\int_{0}^{l_{\max}}\left(  2l+1\right)  \sqrt{\omega_{i,nl}^{2}%
-\frac{l\left(  l+1\right)  }{r^{2}}-m_{i}^{2}\left(  r\right)  }\exp\left(
-\frac{\theta}{4}k_{i}^{2}\right)  \ dl \label{gomega}%
\end{equation}
and the induced cosmological constant becomes%
\[
\frac{\Lambda}{8\pi G}=\frac{1}{6\pi^{2}}\left[  \int_{\sqrt{m_{0}^{2}\left(
r\right)  }}^{+\infty}\sqrt{\left(  \omega^{2}-m_{0}^{2}\left(  r\right)
\right)  ^{3}}e^{-\frac{\theta}{4}\left(  \omega^{2}-m_{0}^{2}\left(
r\right)  \right)  }d\omega\right.
\]%
\begin{equation}
\left.  +\int_{0}^{+\infty}\sqrt{\left(  \omega^{2}+m_{0}^{2}\left(  r\right)
\right)  ^{3}}e^{-\frac{\theta}{4}\left(  \omega^{2}+m_{0}^{2}\left(
r\right)  \right)  }d\omega\right]  , \label{t1loop}%
\end{equation}
where an integration by parts in Eq.$\left(  \ref{tot1loop}\right)  $ has been
done. By further developing the calculations we find%
\begin{equation}
\frac{\Lambda}{8\pi G}=\frac{1}{12\pi^{2}}\left(  \frac{4}{\theta}\right)
^{2}\left(  x\cosh\left(  \frac{x}{2}\right)  -x^{2}\sinh\left(  \frac{x}%
{2}\right)  \right)  \ K_{1}\left(  \frac{x}{2}\right)  +x^{2}\cosh\left(
\frac{x}{2}\right)  K_{0}\left(  \frac{x}{2}\right)  , \label{LambdaNCS}%
\end{equation}
where $K_{0}\left(  x\right)  $ and $K_{1}\left(  x\right)  $ are the modified
Bessel function and%
\begin{equation}
x=\frac{m_{0}^{2}\left(  r\right)  \theta}{4}. \label{xS}%
\end{equation}
The asymptotic properties of $\left(  \ref{LambdaNCS}\right)  $ show that the
one loop contribution is everywhere regular. Indeed, we find that when
$x\rightarrow+\infty$,
\begin{equation}
\frac{\Lambda}{8\pi G}\simeq\frac{1}{6\pi^{2}\theta^{2}}\sqrt{\frac{\pi}{x}%
}\left[  3+\left(  8x^{2}+6x+3\right)  \exp\left(  -x\right)  \right]
\rightarrow0. \label{LNCSz}%
\end{equation}
Conversely, when $x\rightarrow0$, we obtain%
\begin{equation}
\frac{\Lambda}{8\pi G}\simeq\frac{4}{3\pi^{2}\theta^{2}}\left[  2-\left(
\frac{7}{8}+\frac{3}{4}\ln\left(  \frac{x}{4}\right)  +\frac{3}{4}%
\gamma\right)  x^{2}\right]  \rightarrow\frac{8}{3\pi^{2}\theta^{2}}%
\end{equation}
a finite value for $\Lambda$. Note that expression $\left(  \ref{LambdaNCS}%
\right)  $ can be used when the background satisfies the relation
\begin{equation}
m_{0}^{2}\left(  r\right)  =m_{1}^{2}\left(  r\right)  =-m_{2}^{2}\left(
r\right)  . \label{masses1}%
\end{equation}
Examples of metrics satisfying relation $\left(  \ref{masses1}\right)  $ are
the Schwarzschild, Schwarzschild-de Sitter (SdS) and Schwarzschild-Anti de
Sitter (SAdS) backgrounds\footnote{Usually for such geometries, relation
$\left(  \ref{masses1}\right)  $ is satisfied in a region close to the
throat.}. The other interesting cases, namely de Sitter and Anti-de Sitter are
described by%
\begin{equation}
m_{1}^{2}\left(  r\right)  =m_{2}^{2}\left(  r\right)  =m_{0}^{2}\left(
r\right)  , \label{emasses}%
\end{equation}
leading to%
\[
\frac{\Lambda}{8\pi G}=\frac{1}{3\pi^{2}}\left[  \int_{\sqrt{m_{0}^{2}\left(
r\right)  }}^{+\infty}\sqrt{\left(  \omega^{2}-m_{0}^{2}\left(  r\right)
\right)  ^{3}}e^{-\frac{\theta}{4}\left(  \omega^{2}-m_{0}^{2}\left(
r\right)  \right)  }d\omega\right]
\]%
\begin{equation}
=\frac{1}{6\pi^{2}}\left(  \frac{4}{\theta}\right)  ^{2}\left(  \frac{1}%
{2}y\left(  1-y\right)  K_{1}\left(  \frac{y}{2}\right)  +\frac{1}{2}%
y^{2}K_{0}\left(  \frac{y}{2}\right)  \right)  \exp\left(  \frac{y}{2}\right)
. \label{LNCdSAdS}%
\end{equation}
The asymptotic expansion of Eq.$\left(  \ref{LNCdSAdS}\right)  $ leads to%
\begin{equation}
\frac{\Lambda}{8\pi G}\simeq\frac{1}{6\pi^{2}}\left(  \frac{4}{\theta}\right)
^{2}\frac{3}{8}\sqrt{\frac{\pi}{x}}\rightarrow0,
\end{equation}
when $x\rightarrow\infty$. On the other hand, when $x\rightarrow0$, one gets%
\begin{equation}
\frac{\Lambda}{8\pi G}\simeq\frac{1}{6\pi^{2}}\left(  \frac{4}{\theta}\right)
^{2}\left[  1-\frac{x}{2}+\left(  -\frac{7}{16}-\frac{3}{8}\ln\left(  \frac
{x}{4}\right)  -\frac{3}{8}\gamma\right)  x^{2}\right]  \rightarrow\frac
{8}{3\pi^{2}\theta^{2}},
\end{equation}
i.e. a finite value of the cosmological term.

\section{Rainbow's Gravity at work}

In recent years, there has been a proposal on how the fundamental aspects of
special relativity can be modified at very high energies. This modification
has been named \textit{Doubly Special Relativity} (DSR)\cite{GAC}. One of its
effects is that the usual dispersion relation of a massive particle of mass
$m$ is modified into the following expression%
\begin{equation}
E^{2}g_{1}^{2}\left(  E/E_{Pl}\right)  -p^{2}g_{2}^{2}\left(  E/E_{Pl}\right)
=m^{2}, \label{mdisp}%
\end{equation}
where $g_{1}\left(  E/E_{Pl}\right)  $ and $g_{2}\left(  E/E_{Pl}\right)  $
are two functions which have the following property%
\begin{equation}
\lim_{E/E_{Pl}\rightarrow0}g_{1}\left(  E/E_{Pl}\right)  =1\qquad
\text{and}\qquad\lim_{E/E_{Pl}\rightarrow0}g_{2}\left(  E/E_{Pl}\right)  =1.
\end{equation}
Thus, the usual dispersion relation is recovered at low energies. This simple
assumption has a deep impact also when the background is curved. For example,
if we examine the Schwarzschild metric, the analysis of Magueijo and
Smolin\cite{MagSmo} shows that the energy-momentum tensor and the Einstein
equations are replaced by a one parameter family of equations%
\begin{equation}
G_{\mu\nu}\left(  E\right)  =8\pi G\left(  E\right)  T_{\mu\nu}\left(
E\right)  +g_{\mu\nu}\Lambda\left(  E\right)  ,
\end{equation}
where $G\left(  E\right)  $ is an energy dependent Newton's constant, defined
so that $G\left(  0\right)  $ is the physical Newton's constant. Similarly we
have an energy dependent cosmological constant $\Lambda\left(  E\right)  $. A
solution of the \textit{modified Einstein's Field Equations} for a metric of
the form $\left(  \ref{dS}\right)  $ is represented by a \textquotedblleft%
\textit{rainbow metric}\textquotedblright\ whose line element reads%
\begin{equation}
ds^{2}=-\frac{N^{2}\left(  r\right)  dt^{2}}{g_{1}^{2}\left(  E\right)
}+\frac{dr^{2}}{\left(  1-\frac{b\left(  r\right)  }{r}\right)  g_{2}%
^{2}\left(  E\right)  }+\frac{r^{2}}{g_{2}^{2}\left(  E\right)  }\left(
d\theta^{2}+\sin^{2}\theta d\phi^{2}\right)  . \label{line}%
\end{equation}
We expect the functions $g_{1}\left(  E/E_{Pl}\right)  $ and $g_{2}\left(
E/E_{Pl}\right)  $ modify the UV behavior in the same way as GUP and
Noncommutative geometry do, respectively. In presence of Rainbow's Gravity, we
find that Eq.$\left(  \ref{WDW}\right)  $\footnote{Details of the calculation
related to the whole section can be found in Ref.\cite{RGGM}} becomes%
\begin{equation}
\mathcal{\tilde{H}}\Psi=\left[  \frac{g_{1}^{2}\left(  E\right)  }{g_{2}%
^{3}\left(  E\right)  }\tilde{G}_{ijkl}\tilde{\pi}^{ij}\tilde{\pi}%
^{kl}\mathcal{-}\frac{\sqrt{\tilde{g}}}{2\kappa g_{2}\left(  E\right)  }%
\!{}\!\left(  \tilde{R}-\frac{2\Lambda_{c}}{g_{2}^{2}\left(  E\right)
}\right)  \right]  \Psi=0 \label{AccaR}%
\end{equation}
and, consequently Eq.$\left(  \ref{VEV}\right)  $ changes into%
\begin{equation}
\frac{g_{2}^{3}\left(  E\right)  }{\tilde{V}}\frac{\left\langle \Psi\left\vert
\int_{\Sigma}d^{3}x\tilde{\Lambda}_{\Sigma}\right\vert \Psi\right\rangle
}{\left\langle \Psi|\Psi\right\rangle }=-\frac{\Lambda}{\kappa}, \label{WDW3}%
\end{equation}
where%
\begin{equation}
\tilde{\Lambda}_{\Sigma}=\left(  2\kappa\right)  \frac{g_{1}^{2}\left(
E\right)  }{g_{2}^{3}\left(  E\right)  }\tilde{G}_{ijkl}\tilde{\pi}^{ij}%
\tilde{\pi}^{kl}\mathcal{-}\frac{\sqrt{\tilde{g}}\tilde{R}}{\left(
2\kappa\right)  g_{2}\left(  E\right)  }\!{}\!. \label{LambdaR}%
\end{equation}
Of course, Eqs.$\left(  \ref{AccaR},\ref{WDW3}\right)  $ and $\left(
\ref{LambdaR}\right)  $ reduce to the ordinary Eqs.$\left(  \ref{WDW}%
,\ref{VEV}\right)  $ and $\left(  \ref{LambdaSigma}\right)  $ when
$E/E_{Pl}\rightarrow0$. By repeating the procedure of subsection \ref{ps1}, we
find that the TT tensor contribution of Eq.$\left(  \ref{WDW3}\right)  $ is%
\begin{equation}
\hat{\Lambda}_{\Sigma}^{\bot}=\frac{g_{2}^{3}\left(  E\right)  }{4\tilde{V}%
}\int_{\Sigma}d^{3}x\sqrt{\overset{\sim}{\bar{g}}}\tilde{G}^{ijkl}\left[
\left(  2\kappa\right)  \frac{g_{1}^{2}\left(  E\right)  }{g_{2}^{3}\left(
E\right)  }\tilde{K}^{-1\bot}\left(  x,x\right)  _{ijkl}+\frac{1}{\left(
2\kappa\right)  g_{2}\left(  E\right)  }\!{}\left(  \tilde{\bigtriangleup
}_{L\!}^{m}\tilde{K}^{\bot}\left(  x,x\right)  \right)  _{ijkl}\right]
\end{equation}
and the total one loop energy density becomes%
\begin{equation}
\frac{\Lambda}{8\pi G}=-\frac{1}{2}\sum_{\tau}g_{1}\left(  E\right)
g_{2}\left(  E\right)  \left[  \sqrt{E_{1}^{2}\left(  \tau\right)  }%
+\sqrt{E_{2}^{2}\left(  \tau\right)  }\right]  .
\end{equation}
Since Eq.$\left(  \ref{p34}\right)  $ is modified by the \textquotedblleft%
\textit{rainbow metric}\textquotedblright, the two r-dependent radial wave
numbers $\left(  \ref{kTT}\right)  $ become%
\begin{equation}
k_{i}^{2}\left(  r,l,\omega_{i,nl}\right)  =\frac{E_{i,nl}^{2}}{g_{2}%
^{2}\left(  E\right)  }-\frac{l\left(  l+1\right)  }{r^{2}}-m_{i}^{2}\left(
r\right)  \quad i=1,2 \label{kTTr}%
\end{equation}
and with the help of Eqs.$\left(  \ref{p41},\ref{p42}\right)  $, Eq.$\left(
\ref{1loop}\right)  $ reduces to%
\begin{equation}
\frac{\Lambda}{8\pi G}=-\frac{1}{3\pi^{2}}\sum_{i=1}^{2}\int_{E^{\ast}%
}^{+\infty}E_{i}g_{1}\left(  E\right)  g_{2}\left(  E\right)  \frac{d}{dE_{i}%
}\sqrt{\left(  \frac{E_{i}^{2}}{g_{2}^{2}\left(  E\right)  }-m_{i}^{2}\left(
r\right)  \right)  ^{3}}dE_{i}, \label{LoverG}%
\end{equation}
where $E^{\ast}$ is the value which annihilates the argument of the root. In
the previous equation we have assumed that the effective mass does not depend
on the energy $E$. To further proceed, we choose a form of $g_{1}\left(
E/E_{P}\right)  $ and $g_{2}\left(  E/E_{P}\right)  $ which allows a
comparison with the results obtained with a Noncommutative geometry
computation expressed by Eq.$\left(  \ref{t1loop}\right)  $. If we fix%
\begin{equation}
g_{1}\left(  E/E_{P}\right)  =\exp(-\alpha\frac{E^{2}}{E_{P}^{2}}%
)\qquad\text{and}\qquad g_{2}\left(  E/E_{P}\right)  =1, \label{g1g21}%
\end{equation}
with $\alpha>0$, Eq.$\left(  \ref{LoverG}\right)  $ becomes%
\begin{equation}
\frac{\Lambda}{8\pi G}=-\frac{1}{\pi^{2}}\left[  \int_{\sqrt{m_{0}^{2}\left(
r\right)  }}^{+\infty}E^{2}\exp(-\alpha\frac{E^{2}}{E_{P}^{2}})\sqrt{\left(
E^{2}-m_{0}^{2}\left(  r\right)  \right)  }dE\right.
\end{equation}%
\begin{equation}
\left.  +\int_{0}^{+\infty}E^{2}\exp(-\alpha\frac{E^{2}}{E_{P}^{2}}%
)\sqrt{\left(  E^{2}+m_{0}^{2}\left(  r\right)  \right)  }dE\right]  ,
\end{equation}
where we have used relation $\left(  \ref{masses1}\right)  $. Nevertheless we
have to observe that even if the final result is finite and background
dependent, the induced cosmological constant computed with the choice $\left(
\ref{g1g21}\right)  $ will be always negative. When we compare this result
with that obtained in Noncommutative geometry, namely Eq.$\left(
\ref{t1loop}\right)  $, we find that the negativity is principally due the
Rainbow's functions that do not completely enter in the counting of nodes like
in expression $\left(  \ref{gomega}\right)  $. Therefore the pure
\textquotedblleft\textit{Gaussian}\textquotedblright\ choice $\left(
\ref{g1g21}\right)  $ can not give a positive induced cosmological constant.
We are thus led to choose%
\begin{equation}
g_{1}\left(  E/E_{P}\right)  =\left(  1+\beta\frac{E}{E_{P}}\right)
\exp(-\alpha\frac{E^{2}}{E_{P}^{2}})\qquad\text{and}\qquad g_{2}\left(
E/E_{P}\right)  =1, \label{g1g22}%
\end{equation}
with $\alpha>0$ and $\beta\in%
\mathbb{R}
$. Plugging parametrization $\left(  \ref{g1g21}\right)  $ into expression
$\left(  \ref{LoverG}\right)  $, one gets
\[
\frac{\Lambda}{8\pi G}=-\frac{E_{P}^{4}}{4\pi^{2}}\left[  \frac{x^{2}}{\alpha
}\cosh\left(  \frac{\alpha x^{2}}{2}\right)  K_{1}\left(  \frac{\alpha x^{2}%
}{2}\right)  \right.
\]%
\begin{equation}
\left.  -\beta\left(  {\frac{3x}{2{\alpha}^{2}}}-\frac{x^{2}\sqrt{\pi}%
}{{\alpha}^{\frac{3}{2}}}\sinh\left(  \alpha x^{2}\right)  +\frac{3\sqrt{\pi}%
}{2{\alpha}^{\frac{5}{2}}}\cosh\left(  \alpha x^{2}\right)  +\frac{\sqrt{\pi}%
}{2{\alpha}^{\frac{3}{2}}}\left(  x^{2}-\,{\frac{3}{2{\alpha}}}\right)
\,e{^{\alpha x^{2}}}\operatorname{erf}\left(  \sqrt{\alpha\,}x\right)
\right)  \right]  {,} \label{LG}%
\end{equation}
where $x=\sqrt{m_{0}^{2}\left(  r\right)  /E_{P}^{2}}$ and where we have used
expression $\left(  \ref{masses1}\right)  $. The asymptotic expansion for
large $x$ is%
\begin{equation}
\frac{\Lambda}{8\pi G}\simeq-{\frac{\left(  2\beta{\alpha}^{3/2}+\sqrt{\pi
}{\alpha}^{2}\right)  x}{4{\alpha}^{7/2}}-\frac{8\beta{\alpha}^{5/2}%
+3\sqrt{\pi}{\alpha}^{3}}{16{\alpha}^{11/2}x}+\frac{3}{128}}\,{\frac
{16\beta{\alpha}^{7/2}+5\sqrt{\pi}{\alpha}^{4}}{{\alpha}^{15/2}{x}^{3}}%
}+O\left(  x^{-4}\right)  , \label{AsL}%
\end{equation}
while for small $x$, one gets%
\begin{equation}
\frac{\Lambda}{8\pi G}\simeq-{\frac{4{\alpha}^{5/2}+3\sqrt{\pi}\beta{\alpha
}^{2}}{4{\alpha}^{9/2}}}+O\left(  x^{3}\right)  . \label{SmL}%
\end{equation}
If we set%
\begin{equation}
\beta=-{\frac{\sqrt{\alpha\pi}}{2}}, \label{as}%
\end{equation}
then the linear divergent term of the asymptotic expansion $\left(
\ref{AsL}\right)  $ disappears and Eq. $\left(  \ref{LG}\right)  $ vanishes
for large $x$, while for small $x$ we get%
\begin{equation}
\frac{\Lambda}{8\pi G}\simeq{\frac{{3\pi-8}}{8{\alpha}^{2}}}+O\left(
x^{3}\right)  , \label{AsLSm}%
\end{equation}
where we have used the result of expansion $\left(  \ref{SmL}\right)  $. It is
possible to show that with choice $\left(  \ref{as}\right)  $, the induced
cosmological constant is always positive.

\section{Conclusions}

In this contribution we have discussed how modifications of some basic points
of General Relativity lead to finite contributions of ZPE without invoking
neither a regularization nor a renormalization procedure. In particular, we
have considered the effect that a NCG has on the counting of states with a
deep modification of the measure in phase space. Such a modification
introduces a \textit{Gaussian damping} with a natural length: the
Noncommutative length $\theta$. The final result is a finite ZPE, interpreted
as a \textit{Cosmological Constant}, regular at each point of the spacetime.
On the other hand we have MDR's which provide an alternative way to keep under
control UV divergences\cite{RGPLB}. The application of MDR's is done in terms
of a \textquotedblleft\textit{rainbow metric}\textquotedblright\ which,
unfortunately, predicts a \textit{negative cosmological constant} when one
considers the same damping factor suggested from NCG. Nevertheless, by
introducing appropriate variations of the original \textit{Gaussian} proposal,
positive contributions can be obtained.

\bibliographystyle{aipproc}

\end{document}